\documentclass[%
twocolumn,
superscriptaddress,
 amsmath,amssymb,
 aps,
]{revtex4-2}

\usepackage[english]{babel} 
\usepackage[T1]{fontenc}
\usepackage[ansinew]{inputenc}

\usepackage{amsmath}
\usepackage{amsthm}
\usepackage{amsfonts}
\usepackage{bbold}
\usepackage{xcolor}
\usepackage{soul}
\usepackage{dsfont}

\usepackage{graphicx}
\usepackage{dcolumn}
\usepackage{bm}
\usepackage[colorlinks=true,allcolors=blue]{hyperref}
\usepackage{dirtytalk}
\usepackage{braket}
\usepackage{comment}
\newcommand{\B}{\mathbf{B}}

\renewcommand{\r}{\mathbf{r}}
\renewcommand{\H}{\mathbf{H}}
\newcommand{\Ha}{\mathbf{H}_{\rm a}}

\usepackage{siunitx}
\usepackage{upgreek}

\newcommand{\eref}[1]{Eq.~(\ref{#1})}
\newcommand{\fref}[1]{Fig.~\ref{#1}}

\begin{document}

\preprint{v001 - \today}

\title{Magnetic Omniconversion: Source-Independent Molding of Magnetostatic Fields}

\author{Jaume Cunill-Subiranas}
\email{jaume.cunill@uab.cat}
\author{Natanael Bort-Soldevila}
\affiliation{Departament de F\'isica, Universitat Aut\`onoma de Barcelona, 08193 Bellaterra, Barcelona, Spain}
\author{Fabian Resare}
\affiliation{Department of Microtechnology and Nanoscience (MC2), Chalmers University of Technology, SE-412 96 Gothenburg, Sweden}
\author{Nuria Del-Valle}
\affiliation{Departament de F\'isica, Universitat Aut\`onoma de Barcelona, 08193 Bellaterra, Barcelona, Spain}
\author{Witlef Wieczorek}
\email{witlef.wieczorek@chalmers.se}
\affiliation{Department of Microtechnology and Nanoscience (MC2), Chalmers University of Technology, SE-412 96 Gothenburg, Sweden}
\author{Carles Navau}
\email{carles.navau@uab.cat}
\affiliation{Departament de F\'isica, Universitat Aut\`onoma de Barcelona, 08193 Bellaterra, Barcelona, Spain}


\begin{abstract}
Magnetic fields are constrained by the geometry and location of their sources, limiting the ability to freely tailor their spatial distribution. We introduce a general framework to passively convert the magnetic field generated by arbitrary sources into any prescribed desired field within a finite source-free region. Our method relies on field shaping using linear magnetic materials, enabling source-independent magnetic-field molding. We provide the general recipe, analytical and numerical demonstrations for some paradigmatic examples, and a proof-of-concept experiment that validates the idea and materials implementation. This approach enables novel possibilities in magnetic shielding, targeted field delivery, advanced imaging technologies, and a broad range of field-control applications.
\end{abstract}


\maketitle







Precisely generating and controlling magnetic fields in certain regions of space is pivotal in many different scientific and engineering domains. For example, highly uniform magnetic fields are required in particle physics, in medical imaging, for atomic clocks, or in nuclear magnetic resonance \cite{Abel2019_PRA,Albahri2021_PRA,Abel2025_EuroPhysJourC,Cosmus2011_IEEETAS,Gach2020_MedPhys,Sarkany2014_PRL,Hoult1976_ProcRSL}. In precision mass spectroscopy, quantum magnetomechanics, or magnetic confinement fusion, highly precise dipolar, quadrupolar, or more complex fields are essential for the performance of experiments \cite{Gabrielse1999_PRL,Fortagh2007_RMP,Romero-Isart2012_PRL,Bort-Soldevila2024_PRRes,Hofer2023_PRL,Latorre2023_PRAppl,Boozer2005_RevModPhys,Helander2014_RepProgPhys,Bort-Soldevila2024_SciRep}. 

In general, the desired magnetic field distribution in a specified region of space is achieved through a careful selection of magnetic sources that approximate the target field, but never reproduce it exactly. This limitation arises not only from unavoidable imperfections in the experimental implementation, but also because the physical equations used to describe the fields are themselves first-order approximations, or rely on truncated or discretized source representations. Moreover, despite the apparent necessity of superposing multiple coils or magnets to obtain arbitrary field configurations, one could ask whether the field of a single given source can be manipulated to convert it into any desired field in a certain region.

It is well known that magnetic materials can alter field lines by expelling or attracting them, enabling magnetic shielding or field concentration in free space \cite{Sumner1987_JPD,Ferrone2023_Rad,jackson1999classical}. Transformation optics  \cite{Pendry2006_Science,Pendry2012_Science} spurred new methods to control electromagnetic and magnetostatic fields via metamaterials \cite{Kadic2019_NatRevPhys,Magnus2008_NatMat}, enabling perfect lenses or cloaks \cite{Pendry2003_OptExp,Fleury2015_PRAppl,Wood2007_JoP-CM,Gomory2012_Science,Narayana2012_AdvMat,Zhu2015_NatCom}, among other devices \cite{Navau2012_PRL,Bort-Soldevila2024_APL,Barrera2025_ACSnano,Navau2014_PRL,Gargiulo2021_APL, Valadares2024_NatComm,Mach-Batlle2020_PRL}.

Here we show how to generate exactly---and not as an $n$-th order approximation nor as a truncated scheme---any type of magnetostatic field in a finite and bounded source-free region of space using an arbitrary source located outside this region. Our concept is grounded in the idea of magnetic field shaping \cite{Sanchez2021_JAP}, taking it to the next level by transforming, in the finite bounded region, any applied field into any other desired field distribution, which can be, for example, uniform, monopole-like, dipole-like, parabolic-like, etc. As a corollary, our work is a generalization of the concept of magnetic shielding, which aims to nullify the field in a specific region.

In the following, we will state the problem and describe the general solution. We will introduce illustrative and relevant examples, considering the implementation with realistic materials and a proof-of-concept experiment that validates both the general idea as well as the availability of materials with the required properties.

\subsection*{General description of magnetic omniconversion}
We aim to have, in some extended volume, a desired static magnetic field $\H(\r)$, 
independently of the applied sources, which produce a field $\Ha(\r)$. This, which could seem impossible, is indeed possible if the region does not contain the sources of $\Ha$. The idea is to reshape the field lines of the sources by converting them into the desired field lines using adequate magnetic materials surrounding the region of interest.

The desired field $\H(\r)$ must first satisfy $\nabla\cdot\H=0$ and $\nabla\times\H=0$ in a given extended region of space $\mathcal{V}$ in which we assume that there are no sources. One can then write $\H(\r)=-\nabla\phi(\r)$, where $\phi(\r)$ is the scalar magnetic potential. $\phi$ satisfies the Laplace equation $\nabla^2\phi(\r) = 0$ for $\r\in\mathcal{V}$. Because of the uniqueness theorem for the solutions of the Laplace equation, if we find a volume $\mathcal{V}_0 \subset \mathcal{V}$ whose surfaces satisfy the boundary conditions of the desired field, then the magnetic field $\H(\r)$ for $\r\in\mathcal{V}_0$ will have exactly the same shape as the desired one, independently of the field created by the sources $\Ha(\r)$. We now show how to find these surfaces and how to define $\mathcal{V}_0$.

\begin{figure}
    \centering
    \includegraphics[width=1.0\linewidth]{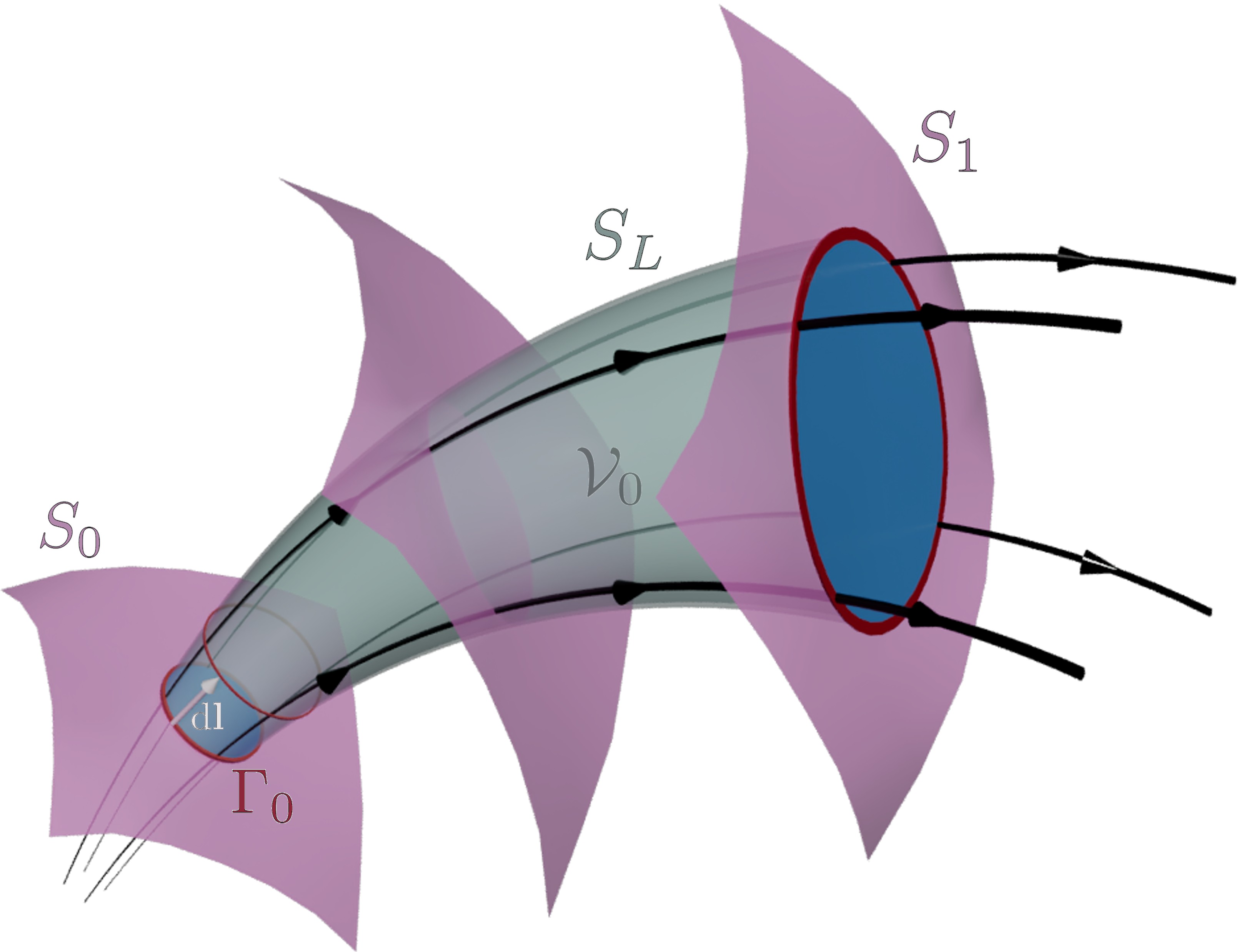}
    \caption{\textbf{Magnetic omniconversion.} Formation of the volume $\mathcal{V}_0$ with the desired field. $S_0$ is an equipotential surface and $\Gamma_0$ any closed line belonging to this surface. Black lines are field lines, and those that thread $S_0$ through $\Gamma_0$ generate a lateral surface, $S_L$. Another equipotential surface, $S_1$, bounds the volume. The white arrow represents a $\text{d}\textbf{l}$ vector of Eq. \eqref{eq:Hparalleldl}.}
    \label{fig:sketch}
\end{figure}

If the field does not have singular points (points where $\H=0$) in the volume $\mathcal{V}$, we start by choosing a closed line $\Gamma_0$ belonging to any equipotential surface $S_0=\{\r \in \mathcal{V} \,| \,\phi(\r)=\phi_0\}$, see \fref{fig:sketch}. In $S_0$, the field $\H=-\nabla\phi$ is perpendicular to the surface. In particular, starting from all the points belonging to $\Gamma_0\subset S_0$, one could generate a new closed line considering differential displacements ${\rm d}{\bf l}$ parallel to the field at each point. That is, the new points are obtained after solving
\begin{equation}
    \H(\r) \times {\rm d}{\bf l}(\r)=0 . 
    \label{eq:Hparalleldl}
\end{equation}
In this way, we can generate a set of closed lines that ultimately form a surface. In fact, Eq. \eqref{eq:Hparalleldl} is a differential equation, whose solution with the $\Gamma_0$ points as boundary conditions yields a surface such that the field $\H$ is parallel to each point of the surface. We call this surface $S_L$.
Finally, we take another equipotential surface $S_1=\{\r\in \mathcal{V} | \phi(\r)=\phi_1 \}$, with $\phi_1\neq\phi_0$. Surfaces $S_0$, $S_1$, and $S_L$ enclose a volume $\mathcal{V}_0\subset \mathcal{V}$ such that in all their boundary surfaces, the field is either parallel (at the points belonging to $S_L$) or perpendicular (at the points belonging to $S_0$ and $S_1$) to the surfaces.  

In other words, we have constructed the volume $\mathcal{V}_0\subset\mathcal{V}$ in such a way that the boundary conditions for the scalar potential $\phi$ are exactly those that yield the desired field $\H$. Because of the uniqueness theorem, the field inside $\mathcal{V}_0$ must have the shape of the desired one. 

To avoid obtaining the trivial solution $\H=0$ for all $\r\in\mathcal{V}_0$, one should locate an applied field source outside the volume $\mathcal{V}_0$ and modify its field lines with adequate materials that satisfy the required conditions at the surfaces $S_0$, $S_1$, and $S_L$. The external field fixes the value of the potentials at $S_0$ and $S_1$. Thus, the inner field's magnitude depends on the external source, but the shape of the field does not. Changing the source (position, nature, magnitude) will change only the magnitude of the magnetic field inside $\mathcal{V}_0$. This will work for \textit{any} external field and \textit{any} desired non-singular magnetic field \footnote{There is an exception: if the field source is located so symmetrically that the potential created in $S_0$ and $S_1$ is exactly the same, then the trivial $\H=0$ solution is the only solution: the field shape remains unchanged, but its magnitude goes to zero.}. We have designed a magnetostatic omniconverter.

For fields containing a singular point, the entire space can be divided into nonintersecting regions such that a field line starting in one region does not cross into another. Additionally, nullclines may appear as lines along which the field is parallel and passes through the singular point (e.g., in cylindrically symmetric systems, the symmetry axis is a nullcline). By arbitrarily setting the singular point at $\phi = 0$, the equipotential surfaces form separate sheets (some with $\phi > 0$, and others with $\phi < 0$). If we want our region $\mathcal{V}_0$ to contain the singular point, the above procedure can be generalized by considering initial closed equipotential lines $\Gamma_0$ that (i) enclose a nullcline, (ii) lie on equipotential sheets with the same sign of $\phi$, and (iii) correspond one-to-one with the separated regions---i.e., there must be as many $\Gamma_0$ lines as there are regions, with each line belonging to a different region. After generating the surfaces $S_L$ from each $\Gamma_0$ using Eq. \eqref{eq:Hparalleldl}, the equipotential surfaces with the other sign of $\phi$ are then used to shut the volume $\mathcal{V}_0$. An example of this process is shown below for a quadrupolar field.

Up to now, the definition of $\mathcal{V}_0$ has been a mathematical construction. We show now how to implement these $S_0$, $S_1$, and $S_L$ surfaces based on physical materials. On the one hand, the equipotential surfaces can be implemented with infinite-magnetic-permeability (IMP) media as materials in which the $\H$-magnetic field is always zero inside them. Equivalently, they can be characterized as linear magnetic materials with permeability $\mu\rightarrow\infty$. 
The tangential component of the field must be continuous across the material surface. For IMP media, this contour condition ensures that the $\H$-field is always perpendicular to the IMP surface, as required for the equipotential surfaces $S_0$ and $S_1$. 
This material is also known as a perfect magnetic conductor \cite{Orton2003_ICAP,Lindell2005_JEWA,Chen2016_JAP}. 
Important to note, soft linear ferromagnetic materials with high permeability (such as mu-metal or cast iron) are excellent practical approximations to the ideal IMP case \cite{Pirottin2025_IEEE-ToM}.

On the other hand, the $S_L$ surface can be implemented using zero-magnetic-permeability (ZMP) materials. These materials have the property that the $\textbf{B}$-field is 0 inside them. Thus, in its exterior surface, the field $\H=\B/\mu_0$ must be parallel to the surface. This is the requirement for the $S_L$ surface.  
Important to note also is that it has been shown in Ref. \cite{Sanchez2021_JAP,Bort-Soldevila2024_SciRep} that superconducting materials (especially type-I ones, with small London penetration depth at the working scales) are good practical approximations to ZMP materials. 

When using superconductors, however, one must consider that ideal superconductors behave as ZMP materials but also exhibit an additional property: they preserve the fluxoid through any closed loop entirely contained within them \cite{Tinkham2004_book}. This imposes an extra constraint. To bypass fluxoid conservation, a possible solution is to introduce a small slit in the surface $S_L$ \cite{Cunill-Subiranas2025_PRB,Navau2014_PRL}, preventing current from circulating around it. In this way, fluxoid conservation no longer applies, and superconductors can function as practical realizations of ZMP materials. Alternatively, if no slit is introduced, superconductors can be field-cooled, meaning the magnetic field present within $\mathcal{V}_0$ after cooling the omniconverter will remain trapped inside, regardless of later changes in the external sources.


\begin{figure*}[t]
    \centering
    \includegraphics[width=1.0\linewidth]{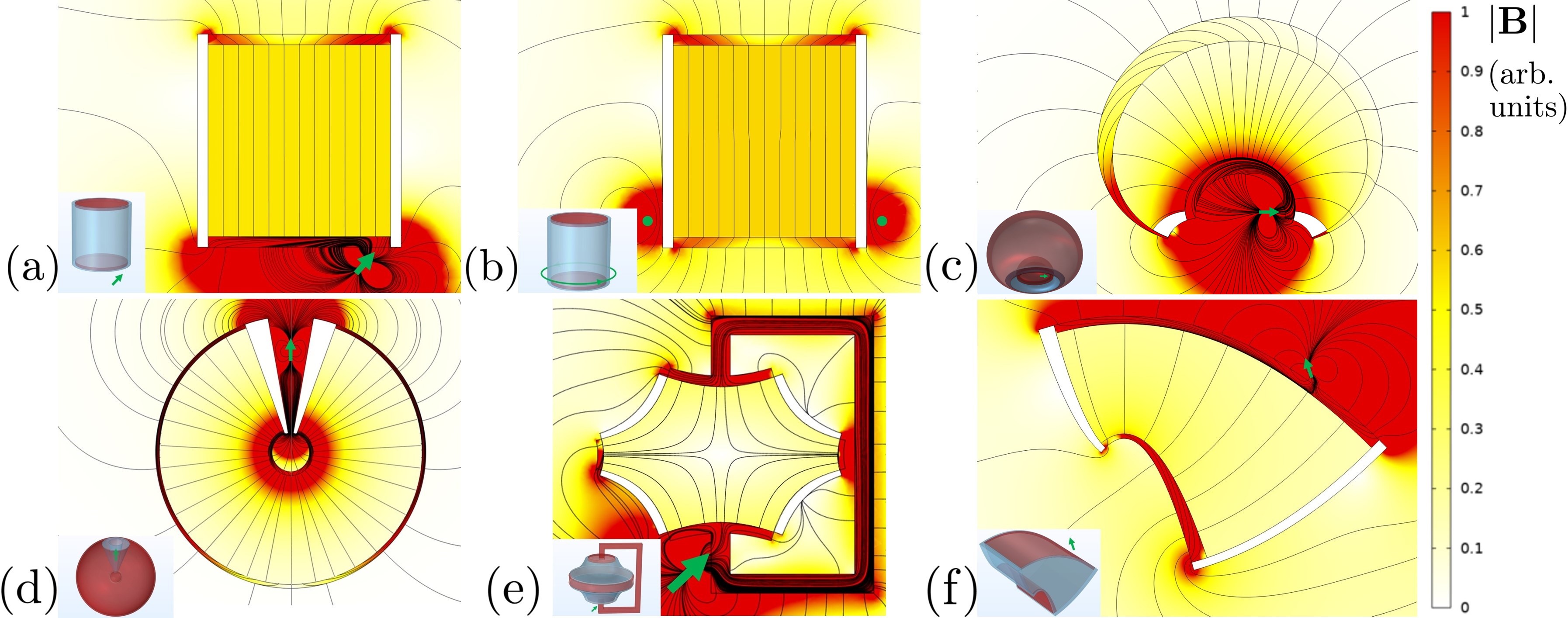}
    \caption{\textbf{Magnetic omniconversion examples.} (a) and (b) A uniform magnetic field is generated within region $\mathcal{V}_0$, regardless of the external source (a dipole in (a), a coil in (b)). (c) The external source is a horizontal dipole, but the desired and generated field is that of a vertical dipole. (d) An external dipole source generates a monopolar field inside $\mathcal{V}_0$. (e) A quadrupolar field with a central singular point is generated. (f) The external source is a dipole, but the generated field lines inside $\mathcal{V}_0$ have a parabolic shape. In the 3D sketches at the bottom left of each case, the IMP surfaces are shown in red and the ZMP surfaces in blue. Green indicates the applied field sources: dipoles in (a, c-f) and a current loop in (b). Note that the magnetic moments of the dipoles are not necessarily equal in magnitude across the different cases, and that the color scale is saturated in some regions for visualization purposes.}
    \label{fig:fields}
\end{figure*}

\subsection*{Representative examples}
In the following, we will explain and formulate in detail how to implement these ideas in some paradigmatic and useful cases. We start with the simplest case of desiring a perfectly uniform field $\H(x,y,z) = H_0 \hat{z}$ in an extended region. 
The scalar potential is $\phi=-H_0 z + C$, where $C$ is an arbitrary constant. An equipotential surface can be the $z=0$ surface and a closed line $\Gamma_0$ can be a circle of radius $R$ in the $z=0$ plane. \eref{eq:Hparalleldl} reduces to $H_0 ({\rm d}y \,\hat{\bf x} - {\rm d}x \,\hat{\bf y})=0$, 
whose obvious solution is ${\rm d}x={\rm d}y=0$. This indicates that, starting with the circle $\Gamma_0$, the surface $S_L$ is a cylinder of radius $R$, parallel to the $z$-axis. The volume can be closed at any other equipotential surface (another plane with constant $z$), say $z=L$. 
This indicates that inside the cylindrical region limited by two IMP materials with internal surfaces at $z=0$ and $z=L$ and a ZMP material with interior surface at $x^2+y^2=R^2$, the magnetic field is exactly uniform, irrespective of the field source one locates outside the cylindrical region (and, of course, outside the materials).
In \fref{fig:fields}(a,b), we show a calculated case where the field in the region of interest is perfectly uniform, regardless of whether the field source is a dipole or a coil.

Following the same principle, one can also generate an exact dipolar field in a given region, irrespective of the source fields.  In this case, the desired field and the corresponding scalar potential, using standard spherical coordinates $(r,\theta,\varphi)$ and, without loss of generality, assuming that the desired dipolar field is that of a dipole at the origin with magnetic moment ${\bf m}=m \hat{{\bf z}}$, are
$\H (r,\theta,\varphi) = \frac{m}{4\pi r^3}\left( 2\cos\theta \hat{{\bf r}}+ \sin\theta \hat{\boldsymbol{\uptheta}} \right)$
and $
\phi (r,\theta,\varphi) = \frac{m}{4\pi} \frac{\cos\theta }{r^2}$, respectively.
An equipotential surface with scalar potential $\phi_0$ is defined by those points satisfying 
\begin{equation}
r^2= A \cos\theta,
\end{equation} 
for a given $A=\frac{m}{4\pi\phi_0}$. From any closed line $\Gamma_0$ on this surface, the lateral $S_L$ surface is found by solving Eq. \eqref{eq:Hparalleldl} which in this case becomes
$H_r r \, {\rm d}\theta = H_\theta \, {\rm d}r.$
For $r\neq 0$, the solution is 
\begin{equation}
r=C \sin^2\theta,
\end{equation}
where $C$ is a constant determined by the specific $\Gamma_0$ selected. 
It's crucial to note that the region where the field is exactly dipolar must avoid the $\r=0$ point, as the external sources must be outside the $\mathcal{V}_0$ region. As shown in \fref{fig:fields}(c), we have created an exact dipolar field as if it were generated by a vertical dipole at the origin, but which is actually generated by a horizontal dipole outside the region of interest. Hence, this confirms once more that the field shape inside the region $\mathcal{V}_0$ is independent of the field sources outside that region.

Another interesting case is to have exactly a monopolar field in a given region. Approximate monopolar fields can be obtained from superconducting vortices \cite{Clem1991_PRB}, for example. Here, however, we demonstrate that an exact monopolar field can be generated in a large region of space from any external source. In this case, the desired field and scalar potential have the form (in spherical coordinates)
$ \H(\r) =\frac{A}{r^2} {\bf \hat{r}}$ and $\phi(\r) = -\frac{A}{r} + C$, respectively,
being $A$ and $C$ constants. To define the volume $\mathcal{V}_0$, we start with a sphere of radius $R_0$ centered at the origin (equipotential surface) and a closed line $\Gamma_0$ as the intersection of this sphere with the plane $z=z_0$ for $z_0<R_0$. Choosing the surface of the sphere with $z<z_0$, we construct the volume by finding the ${\rm d}{\bf l}$ using \eref{eq:Hparalleldl}. This reveals that the new surface, $S_L$, is simply constructed by increasing the radius (as the solution of \eref{eq:Hparalleldl} is ${\rm d}\theta={\rm d}\varphi = 0$). To enclose the volume, we select another sphere of radius $R_1>R_0$. This volume and the simulation using a dipole as the external source are shown in \fref{fig:fields}(d).

If we want to generate an exact linear quadrupolar field and include the singular point within $\mathcal{V}_0$, we must consider several starting $\Gamma_0$ lines, as explained previously. This field and the scalar potential can be expressed as: $\H(x,y,z) = A_0 (-x \hat{\bf x} - y\hat{\bf y} +2 z \hat{\bf z})$, 
$\phi(x,y,z) = A_0 \left( -\frac{x^2}{2}-\frac{y^2}{2}+z^2\right)$, for some constant $A_0$. Without loss of generality, we have assumed the potential at the singular point ($\r=0$) is zero, that $A_0>0$, and that the quadrupolar field is symmetric with respect to the $z$-axis. In this case, there are several disconnected sheets of equipotential surfaces, defined by
\begin{equation}
    z^2 = \frac{\phi_0}{A_0} + \frac{\rho^2}{2}\, .
\end{equation}
For $\phi_0>0$, it represents two sheets of surfaces lying on the $|z|>\rho/2$ volume (one for $z>0$ and one for $z<0$). For $\phi_0<0$, it describes a cylindrically symmetric surface lying in the region $\rho>2\, |z|$. Additionally, no field line can cross the $z=0$ plane (2D manifold), effectively dividing the space into two regions: $z>0$ and $z<0$. To define the volume $\mathcal{V}_0$, one should consider two equipotential closed lines, $\Gamma_{01}$ and $\Gamma_{02}$, 
one in the $z>0$ region and the other in the $z<0$ region and both circulating the $z$-axis (1D nullcline). From both lines, one builds the lateral surfaces $S_{L1}$ and $S_{L2}$ using Eq. \eqref{eq:Hparalleldl}. Lateral surfaces satisfy 
\begin{equation}
\rho z^2 = C_i\, ,
\end{equation}
with $i\in\{1,2\}$. $C_1$ and $C_2$ are determined by the selection of $\Gamma_{01}$ and $\Gamma_{02}$ lines, respectively. Note that each $\Gamma_0$ yields one $S_L$ surface. Finally, the volume $\mathcal{V}_0$ is determined by closing the volume with another equipotential surface with a potential of opposite sign. A simulation illustrating the transformation of a dipole field into a quadrupolar field containing the singular point is shown in \fref{fig:fields}(e). Note that, to make sure the two IMP surfaces actually have the same potential as intended, they need to be magnetically connected using IMP material outside $\mathcal{V}_0$.

If we want to achieve a particular geometry for the field lines instead of a specific expression for the field, the ideas discussed above can be applied in a straight-forward way. One would build a ZMP surface with the geometry of the field lines and find a set of surfaces perpendicular to the field, choose two IMP of such surfaces that, together with the ZMP surface, enclose a volume $\mathcal{V}_0$ where the field will have exactly the desired shape. For example, suppose one desires a field with parabolic field lines. In that case, one can choose a ZMP surface in the shape of an extruded parabola and use the corresponding conjugate-parabola surfaces as IMP surfaces. Then, any source located outside the volume $\mathcal{V}_0$ will generate, inside $\mathcal{V}_0$, a field with parabolic field lines, as desired. This example is shown in \fref{fig:fields}(f).

If we want to generate a given field in a non-simply connected volume (i.e., a volume with a hole \cite{Albahri2021_PRA}), we can choose, for the non-singular case, two different non-intersecting starting  lines $\Gamma_0$ with the same potential. From each of these, we construct the lateral surfaces. Once enclosed by another equipotential surface $S_1$, the resulting $\mathcal{V}_0$ volume will be non-simply connected and, by construction, the field inside it will have the desired shape. More complicated cases can also be envisaged as combinations of the presented ideas. Another particular case would be that $S_L$ already determines a closed volume (i.e., a toroid). In that case, there would be no need for IMP surfaces, but one should use an external source that ensures that the field is not zero inside \cite{Bort-Soldevila2024_SciRep}.

\begin{figure*}[t]
    \centering
    \includegraphics[width=1.0\linewidth]{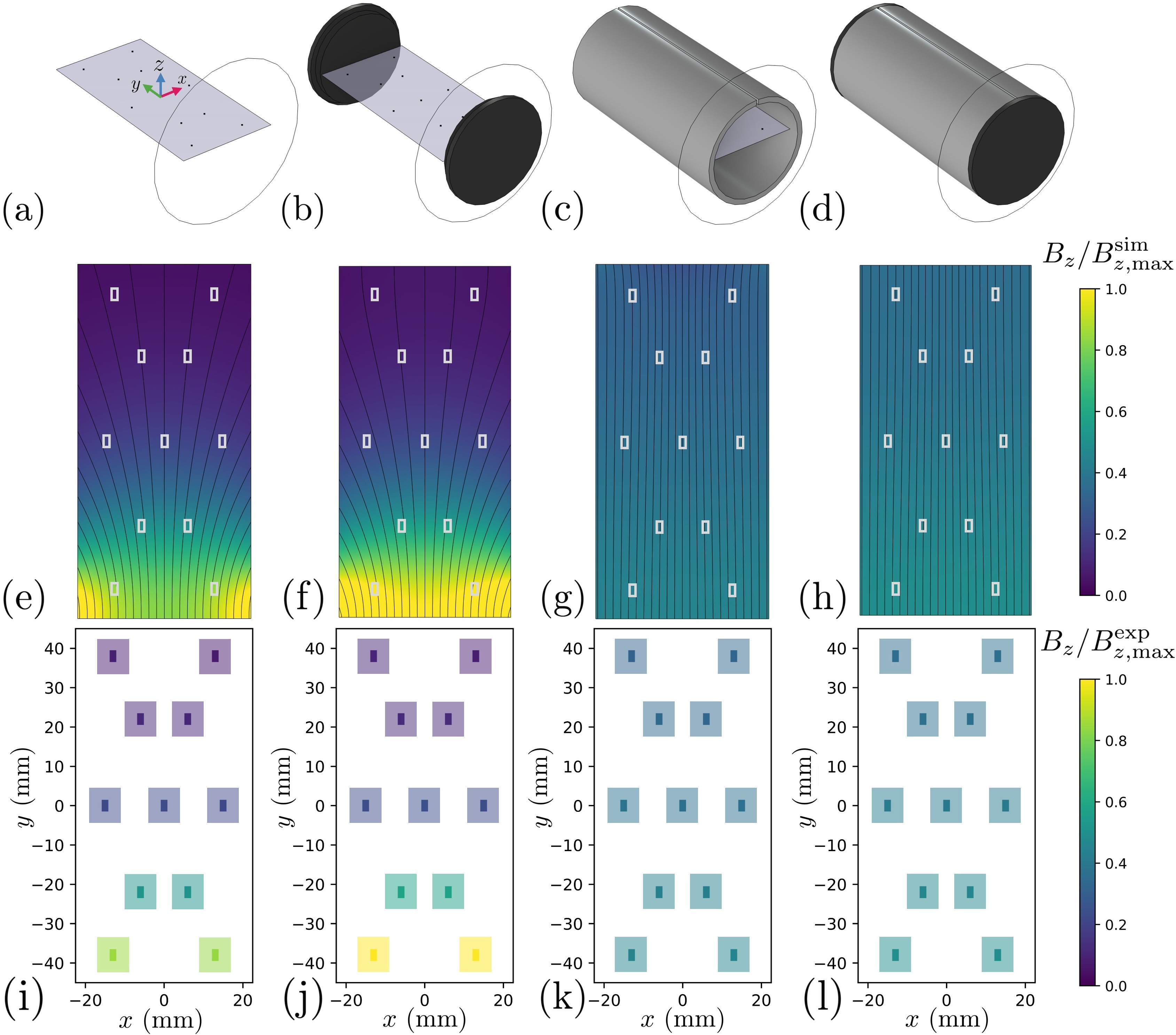}
    \caption{\textbf{Proof-of-concept magnetic omniconversion demonstrator.} Experimental system configurations (a)-(d), corresponding simulation results (e)-(h), and experimental measurements (i)-(l). The induced magnetic fields from the same flat circular coil are compared under four configurations: (a), (e), and (i) coil and PCB with sensors only; (b), (f), and (j) with added soft ferromagnetic caps; (c), (g), and (k) with the superconducting aluminum tube added but without the caps; and (d), (h), and (l) with the full assembly-tube and caps. Color scales represent the axial magnetic field component, $B_z$, normalized to the respective maximum values in simulation and experiment, $B_{z,\text{max}}^\text{sim}$ and $B_{z,\text{max}}^\text{exp}$. In the simulations (e)-(h), black lines indicate magnetic field lines, and white rectangles mark the actual sensor positions. The same rectangles are used in the experimental plots (i)-(l), uniformly colored according to the $B_z$ value measured by each sensor, and framed by a larger rectangle of the same color with reduced opacity to enhance visibility.}
    \label{fig:experiment}
\end{figure*}

\subsection*{Proof-of-principle experiment}
In the following, we illustrate the basic working principle of our proposed method through a proof-of-concept experiment. To this end, we fabricated a proof-of-concept device---an omniconverter---designed to transform the non-uniform field of a single planar circular coil into a homogeneous field within a defined spatial region. This study region is a 92 mm long cylinder with a 22.5 mm radius, located externally to the coil (as in the example of \fref{fig:fields}(b)). The cylinder's end caps, made of cast iron, act as an IMP medium, oriented perpendicular to the desired magnetic field. The cylindrical side wall is a 100 mm long, 2.5 mm thick tube of high-purity aluminum, which is a type-I superconductor at our working temperatures ($<150$ mK).  The London penetration depth is $\simeq 50$ nm, negligible with respect to all the dimensions of the tube. Thus, the aluminum acts as a ZMP material aligned parallel to the field lines.  A 1 mm axial slit in the aluminum tube oriented in the axial direction prevents net current circulation (as the system is zero-field cooled). 

We mapped the magnetic field within the study region using 11 commercial tunneling magnetoresistance sensors, soldered onto a custom printed circuit board (PCB) and aligned axially inside the tube [\fref{fig:experiment}(a-d)], forming a pattern that enables the study of the field's homogeneity. The entire system was housed in a dilution refrigerator, with a multiturn coil wound from NbTi wire coaxially positioned 12 mm from the aluminum tube to prevent the tube from exceeding its critical magnetic field.

Measurements were conducted across four distinct configurations to compare the magnetic field produced by: (a) the coil alone, (b) the coil with cast-iron end caps, (c) the coil with the superconducting aluminum tube, and (d) the complete omniconverter system. The aluminum tube and PCB remained fixed, while the cast-iron caps were only installed for cases (b) and (d). The superconducting state of the aluminum tube was controlled by temperature, activating below 1.0 K [cases (c) and (d)] and deactivating above this temperature [cases (a) and (b)]. For each case, current was swept through the coil from -0.1 A to 0.1 A. 
Room temperature data for cases (a) and (b) were corrected for thermal drift in sensor outputs using measurements at 1.28 K, accounting for sensor response variations at cryogenic temperatures.
Given the linear response of the sensors, the measured voltage was directly mapped to the axial component of the magnetic field via a linear fit. Details are provided in the Supplementary Information. These normalized axial magnetic field measurements are presented in \fref{fig:experiment}(i-l).

For comparison, four corresponding numerical simulations were performed using the finite element method to solve Amp\`ere's law. Cast-iron caps were modeled as a nearly ideal IMP ($\mu/\mu_0 = 10^5$) material, and superconducting aluminum as an almost ideal ZMP ($\mu/\mu_0 = 10^{-5}$) medium. Simulated normalized axial field components are shown in \fref{fig:experiment}(e-h).

A clear agreement is observed between the experimental [\fref{fig:experiment}(i-l)] and simulated [\fref{fig:experiment}(e-h)] relative magnetic field values. The standalone coil generates a non-uniform field [\fref{fig:experiment}(i)], confirmed by the curved field lines in simulation [\fref{fig:experiment}(e)].

With the cast-iron end caps, the field remains non-uniform, showing enhancement near the cap closest to the coil [\fref{fig:experiment}(j)]. However, simulations [\fref{fig:experiment}(f)] indicate a slight improvement in field uniformity towards the tube ends.

Crucially, when the aluminum tube becomes superconducting, a distinct and dramatic change occurs: the magnetic field inside the tube becomes \textit{uniformly distributed}, with field lines transforming into straight, parallel lines perpendicular to the tube's cross-section [\fref{fig:experiment}(k,l)]. Given the tube's aspect ratio, the superconductor's influence on field uniformity is more dominant than that of the ferromagnetic caps alone, explaining the observed similarities between cases (g,k) and (h,l). However, the influence of the ferromagnetic material would become more dominant closer to the edges or if the tube were wider; for instance, a slight bending of field lines is visible in the upper region of \fref{fig:experiment}(g), absent in \fref{fig:experiment}(h) due to the presence of the caps.

In the fully assembled omniconverter configuration [\fref{fig:experiment}(d,h,l)], remaining minor deviations from perfect field uniformity are not primarily due to real material imperfections. This is supported by the close agreement between simulation and experiment, \fref{fig:experiment}(h) and \fref{fig:experiment}(l), indicating that the actual materials faithfully implement ideal IMP and ZMP behavior at these scales. Instead, these non-uniformities are attributed to the presence of the slit, which is also accounted for in the simulations. This confirms the suitability of the chosen materials for practical implementation and suggests that further improvements should focus on minimizing the slit separation.

\subsection*{Conclusions}
In this work, we have focused on magnetostatic fields, as IMP and ZMP surfaces can be effectively approximated by conventional materials such as cast iron and superconducting aluminum, respectively. However, the same theoretical principles could be extended to electrostatic fields 
if one could find materials that approximate infinite-electrical-permittivity---ideal conductors---and statically zero-electrical-permittivity media. 
The procedure presented in this work is a general recipe to generate the desired field with the available sources, so it is relevant far beyond a particular application. In principle, one can envisage any applied field and, designing the adequate omniconverter, convert any source field into the desired one. This could be beneficial for improving existing technologies---such as particle traps, medical imaging, or particle accelerators---by obtaining more controllable magnetic fields. Moreover, the present work paves the way for innovative solutions that would otherwise appear unattainable if they relied on inaccessible magnetic field configurations.

\begin{acknowledgments} 
We thank Lars J$\ddot{\text{o}}$nsson for his assistance in fabricating the mechanical parts of the experiment, and Samuel Lara-Avila and Thilo Bauch for their valuable advice on the magnetic sensors.
This work was supported in part by the Horizon Europe 2021-2027 Framework Programme (European Union) through the SuperMeQ project (Grant Agreement number 101080143) and by Spanish project PID2023-149054NB-I00 from MCIU/AEI/10.13039/501100011033/FEDER,UE. J.C.-S. acknowledges funding from AGAUR-FI Joan Or\'o grants (Grant No. 2025 FI-3 00143) of the Generalitat de Catalunya and the European Social Fund Plus. W.W.~acknowledges support by the European Research Council under Grant No. 101087847 (ERC Consolidator SuperQLev) and the Knut and Alice Wallenberg (KAW) Foundation through a Wallenberg Academy Fellowship and Scholar. 
\end{acknowledgments}



\vspace{12pt}

\bibliographystyle{apsrev4-2}

\bibliography{Bib_exactfields} 

\end{document}


\maketitle
{\scriptsize
\noindent
\textit{$^1$ Departament de F\'isica, Universitat Aut\`onoma de Barcelona, 08193 Bellaterra, Barcelona, Spain}

\noindent
\textit{$^2$ Department of Microtechnology and Nanoscience (MC2), Chalmers University of Technology, SE-412 96 Gothenburg, Sweden}
}

\noindent\rule{2cm}{0.4pt}

{\footnotesize $^{\text{a)}}$ Electronic mail: jaume.cunill@uab.cat}

{\footnotesize $^{\text{b)}}$ Electronic mail: witlef.wieczorek@chalmers.se}

{\footnotesize $^{\text{c)}}$ Electronic mail: carles.navau@uab.cat}

\section{Experimental details}

\subsection{The system}

\begin{figure}
    \centering
    \includegraphics[width=0.75\linewidth]{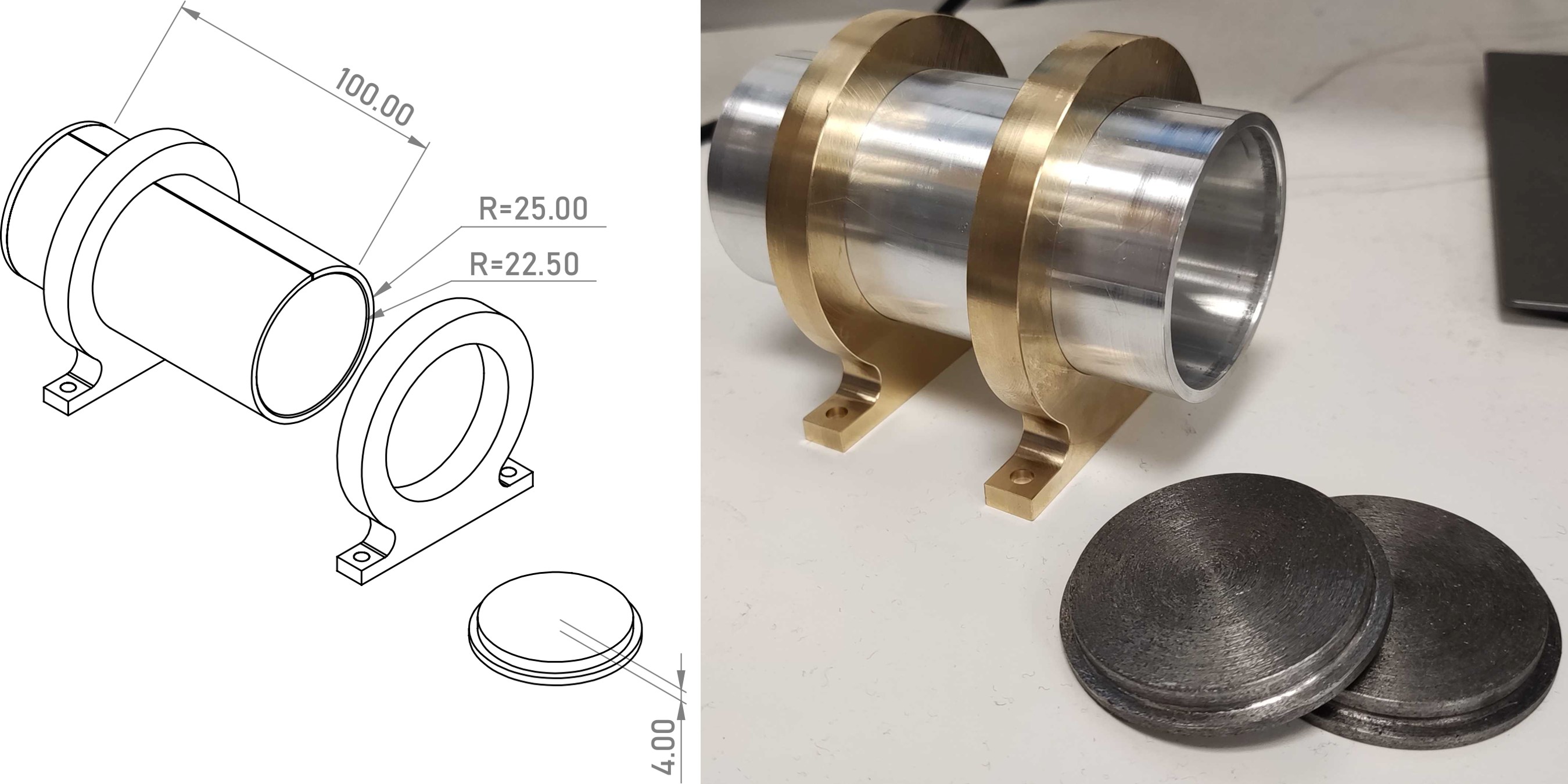}
    \caption{\textbf{Geometry of the aluminum tube assembly.} Dimensions (in mm, left) and photograph (right) of the aluminum tube with two cast-iron end caps and two brass supports.}
    \label{fig:system}
\end{figure}

The experimental setup consists of a high-purity aluminum tube (4N) with a wall thickness of 2.5 mm, an inner radius of 22.5 mm, and a length of 100 mm (see \fref{fig:system}). The end caps are made of cast iron, each with a thickness of 6.5 mm, of which 4 mm are inserted inside the tube and 2.5 mm extend longitudinally. Throughout the experiment, the aluminum tube is positioned horizontally inside the dilution-refrigerator cryostat (Bluefors LD250), supported by press-fitting it into two brass mounts and secured with screws to the mixing stage.

\begin{figure}
    \centering
    \includegraphics[width=1.0\linewidth]{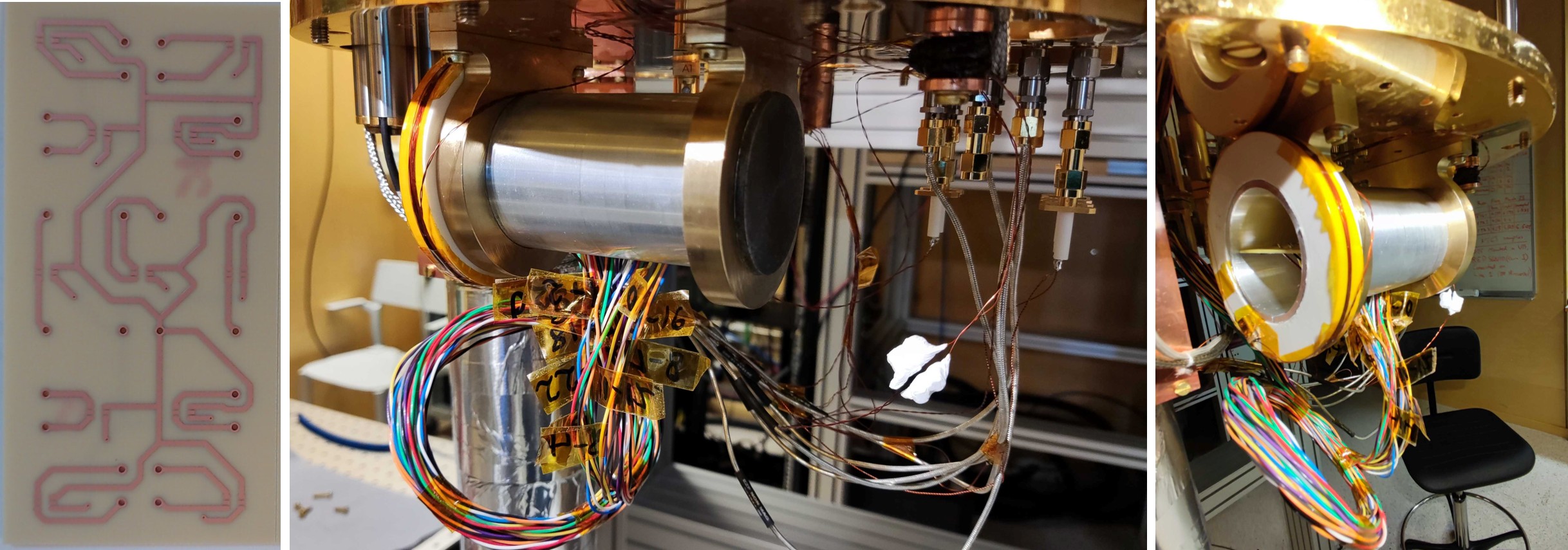}
    \caption{\textbf{Photographs of the PCB and final assembly within the cryostat.} Photographs of the PCB before soldering the commercial Allegro CT100 sensors (left), and the complete experimental assembly integrated into the cryostat, shown with caps (center) and without caps (right).}
    \label{fig:pcb-photos}
\end{figure}

The printed circuit board (PCB) used for measurements contains 11 commercial 1D linear tunneling magnetoresistance (TMR) sensors, Allegro CT100, arranged to map the axial component of the field across different regions of the tube. 
In \fref{fig:pcb-photos} (left), the two central-bottom ports are used to connect the supply voltage cables for all sensors, while the remaining 22 cables are used to record measurements depending on the sensor being monitored. The sensors were soldered onto the PCB using lead-free solder paste.


The board is placed horizontally inside the tube, with its dimensions (90 mm $\times$ 45 mm) such that the tube itself acts as a support, securing it in place to prevent movement. The aluminum tube has a 1 mm thick slit running the entire length, parallel to the tube axis. Through this slit, the 24 copper wires pass.

A multi-turn NbTi coil is used as the magnetic field source. It is mounted near the aluminum tube on a 3D printed spool, leaving a 12 mm gap between the coil and the tube. This spacing prevents exceeding the critical field of aluminum, which might occur if the coil were wound directly onto the tube. Throughout the experiments, the coil is always positioned in the same location, near one end of the tube (adjacent to one of the brass mounts, as in \fref{fig:pcb-photos}).

\subsection{Data processing}

\subsubsection{Raw data}

To carry out the measurements of the magnetostatic field, the voltage values measured by each of the 11 sensors on the PCB were recorded as a function of the current flowing through the NbTi coil. This procedure was repeated under different conditions in order to obtain the four distinct cases presented in Fig. 3 of the main article.

\begin{figure}
    \centering
    \includegraphics[width=1.0\linewidth]{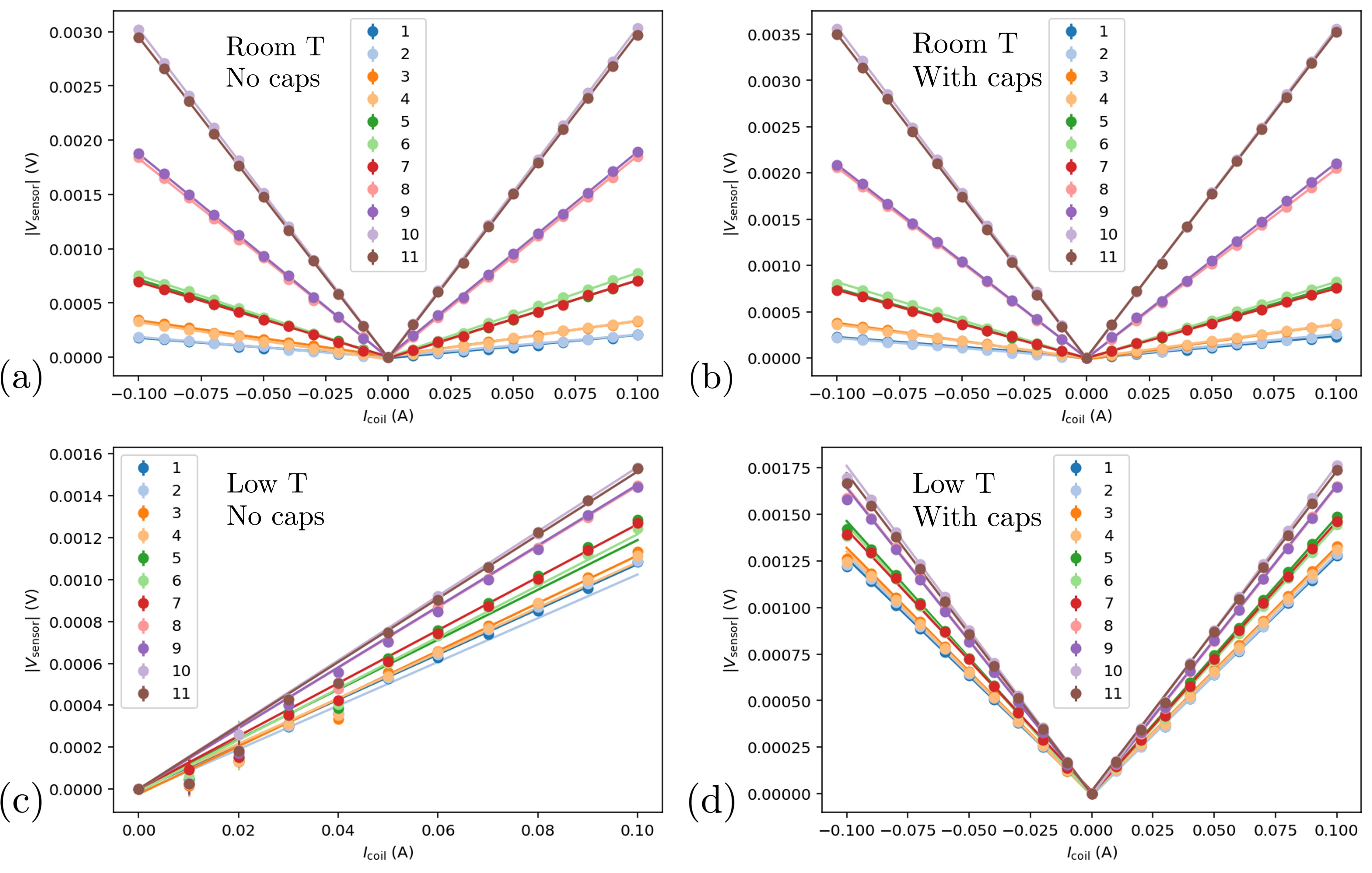}
    \caption{\textbf{Experimental measurements.} Absolute values of voltage measured at each of the 11 TMR sensors as a function of the current flowing through the NbTi coil. Each data point represents the mean of five consecutive voltage measurements. Solid lines are linear fits and, for most points, error bars are smaller than the symbol size. (a) System at room temperature and the cast-iron caps are not in place. (b) System at room temperature and the cast-iron caps are mounted on the tube ends. (c) System below 1.0 K and the cast-iron caps are not in place. (d) System below 1.0 K and the cast-iron caps are mounted on the tube ends.}
    \label{fig:measurements}
\end{figure}

In \fref{fig:measurements}(a), we show the absolute values of the recorded voltages when the system is at room temperature and the cast-iron caps are not in place. Each data point represents the mean of five consecutive voltage measurements. The numbering of the sensors follows the order from left to right and from top to bottom, as shown in Fig. 3. For the case shown in \fref{fig:measurements}(b), where the caps are mounted on the tube ends at room temperature, we also plot the corresponding average voltages.



This same system was measured at temperatures below 1.0 K (zero-field cooling), allowing the high-purity aluminum tube to enter the superconducting state. Although the sample itself lacked a temperature sensor, we monitored the temperature using the mixing-stage sensor of the cryostat. 
The results are shown in \fref{fig:measurements}(c) without the cast-iron caps and in \fref{fig:measurements}(d) with the caps present.



However, since the temperature difference between the measurements performed at room temperature and those at cryogenic temperatures is large and slightly affects the sensors, the voltages were also recorded at a cryogenic temperature of ($1.28\pm0.01$) K, which is above the critical temperature of aluminum. These measured values are shown in \fref{fig:temperatures}(a) and allow a correction factor to be applied to the voltages measured at room temperature for comparison with those obtained at low temperatures.

\subsubsection{Linear fits}

By performing a linear fit of the sensor voltages as a function of the coil current (which is proportional to the applied field), the slopes of each linear regression can be obtained. These slopes are directly proportional to the field experienced by each sensor relative to the applied field. The slopes resulting from these linear regressions are normalized and shown in \tref{tab:data}. All correlation coefficients from the fits range between 0.996 and 0.999990. Voltages measured at room temperature were corrected by multiplying each sensor's values by the mean of the ratios between its voltages at each coil current in \fref{fig:temperatures}(a) and the corresponding voltages in \fref{fig:measurements}(a), allowing direct comparison with the cryogenic measurements. These mean ratios (i.e., correction factors) range from 0.97 for sensor 9 to 1.62 for sensor 1, indicating that although the effect of such large temperature changes on the sensors is noticeable, it remains moderate. 

\begin{table}
\centering
\caption{\textbf{Normalized magnetic field results.} Values of $B_z/B_{z,\text{max}}^\text{exp}$, derived from linear-fit slopes obtained from the data points in \fref{fig:measurements}(a-d), respectively. A correction factor derived from data in \fref{fig:temperatures}(a) was applied to the (*) cases for proper comparison at cryogenic temperatures. Uncertainty on all values: $\pm0.010$.}
\label{tab:data}
\begin{tabular}{c||c|c|c|c}
\hline \hline
Sensor \# & Coil* & Coil+Caps* & Coil+Tube & Coil+Tube+Caps \\\hline
1 & $0.086$ & $0.105$ & $0.316$ & $0.357$\\
2 & $0.071$ & $0.084$ & $0.324$ & $0.360$\\
3 & $0.110$ & $0.118$ & $0.336$ & $0.371$\\
4 & $0.106$ & $0.119$ & $0.331$ & $0.367$\\
5 & $0.219$ & $0.236$ & $0.380$ & $0.415$\\
6 & $0.228$ & $0.243$ & $0.366$ & $0.404$\\
7 & $0.217$ & $0.232$ & $0.369$ & $0.408$\\
8 & $0.523$ & $0.584$ & $0.430$ & $0.463$\\
9 & $0.519$ & $0.578$ & $0.428$ & $0.461$\\
10 & $0.849$ & $1.000$ & $0.451$ & $0.494$\\
11 & $0.840$ & $0.996$ & $0.456$ & $0.486$\\
\hline \hline
\end{tabular}
\end{table}

\subsection{Transition to superconductivity}

\begin{figure}
    \centering
    \includegraphics[width=1.0\linewidth]{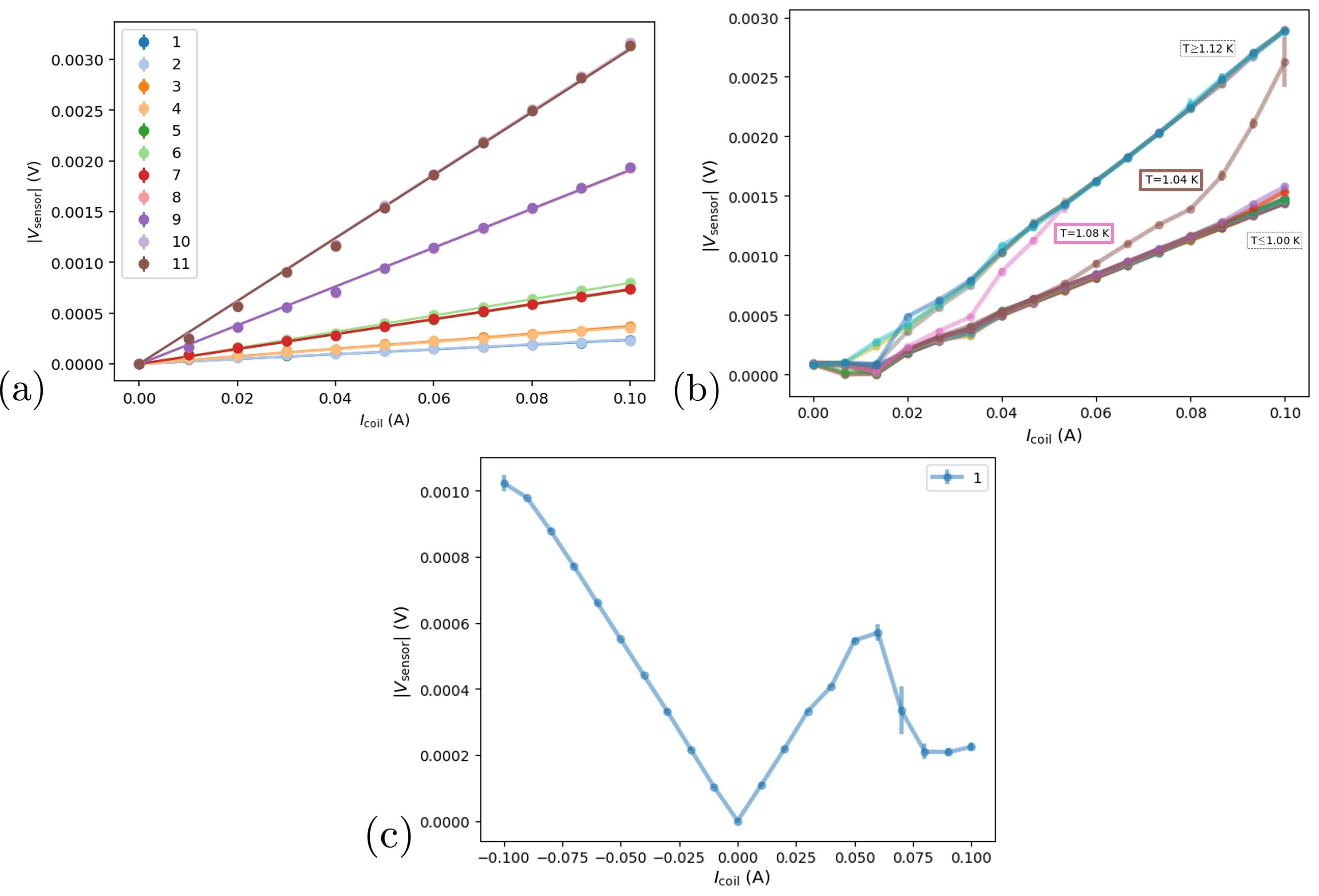}
    \caption{\textbf{Temperature effects while measuring.} Absolute values of voltage measured for some of the TMR sensors as a function of the current flowing through the NbTi coil. Measurements were performed without the cast-iron caps in the tube. Each point represents the mean of five consecutive measurements [with a 0.25 s and 0.50 s interval between them in panels (a,b) and (c), respectively]. For most points, error bars are smaller than the symbol size.
    (a) Voltages measured by the 11 TMR sensors at a cryogenic temperature of (1.28$\pm$0.02) K, above the critical temperature of aluminum. Solid lines are linear fits. (b) Voltage measured by sensor number 11 over a temperature range from 0.04 K to 1.24 K in 0.04 K increments. Solid lines are guides for the eye. Uncertainty on all temperatures: $\pm$0.02 K. The overlapping upper lines correspond to temperatures between 1.12 K and 1.24 K, while the overlapping lower lines correspond to temperatures between 0.04 K and 1.00 K. In these ranges, no change in magnetic field is observed, showing that the aluminum is superconducting below 1.00 K and in the normal state above 1.12 K. In contrast, measurements at 1.04 K and 1.08 K show a change in voltage slope, evidencing the superconducting transition. (c) Voltage measured by sensor number 1 \textit{starting} below the critical temperature of aluminum ($<$1 K). Solid lines are guides for the eye. The measurement was performed without temperature control during acquisition.}
    \label{fig:temperatures}
\end{figure}

In this experiment, the critical temperature of high-purity (4N) aluminum was also measured indirectly, using the TMR sensors to detect the differences in the magnetic field when the tube is superconducting or not. 
To this end, the configuration with the PCB and sensors placed inside the aluminum tube, but without the cast-iron caps, was used. The voltage recorded by sensor number 11 was measured while driving currents 
through the superconducting coil. 
These measurements were repeated over a temperature range from 0.04 K to 1.24 K, in increments of 0.04 K. The results are shown in \fref{fig:temperatures}(b).

Up to 1.00 K, no significant changes are observed in the slopes of the voltage-versus-current curves for sensor 11. However, between 1.00 K and 1.12 K, the curves show a clear increase in the magnetic field at the position of sensor 11 (located near the coil). Above 1.12 K, the linear fits no longer exhibit noticeable changes, while maintaining a slope clearly higher than that observed below 1.00 K. Since sensor 11 is located near the coil, when the aluminum becomes superconducting, the magnetic field decreases in that region, as it redistributes more uniformly along the tube. However, once the critical temperature of the material is exceeded and it ceases to behave diamagnetically, the field at positions close to the NbTi coil increases. Because the temperature was monitored using the mixing-stage sensor rather than a sensor on the sample---and thus does not directly measure the material's temperature---our data constrain the superconducting transition of this 4N bulk aluminum to the range 1.00-1.12 K.

This phenomenon is relevant for the experiment, since the PCB connections are not superconducting. Consequently, when performing a current ramp, the sensors heat up and increase the temperature of the whole system. However, the ramp can be carried out quickly enough to avoid breaking the superconductivity of the aluminum tube within a single measurement. If this is not taken into account and the measurement is performed slowly while maintaining a continuous bias current through the PCB, field profiles such as that shown for sensor 1 in \fref{fig:temperatures}(c) can be obtained: the tube is initially superconducting, but above 50 mA, due to heating accumulated during previous measurements, a sudden change in slope (i.e., magnetic field) is observed as the critical temperature is exceeded. 


\subsection{About the TMR sensors}


The sensors soldered onto the PCB and used to measure the magnetostatic field inside the tube are commercial devices: Allegro, 
model CT100 (6-lead SOT-23 package).

Each of these sensors measures a single component of the magnetic field. In this experiment, to simplify the demonstration, all sensors were oriented along the axial direction of the tube, since this is the direction of the uniform field to be generated, and the component of the magnetic field in this direction is expected to be the largest. The layout (pattern) of the sensors on the PCB was chosen to provide representative measurements of the field uniformity inside the tube and along its axial direction, using a total of 11 sensors. To obtain a more complete mapping and increase the spatial resolution, a larger number of sensors would be required.

Throughout the entire experiment, each sensor was supplied with a voltage of 3 V, which lies within the operating range recommended by the manufacturer (between 1.0 V and 5.5 V). The magnitude of the applied magnetic fields is also well within the linear operating range of the sensors. According to the datasheet, they exhibit linearity over a range of $\pm20$ mT, with an associated error of $\pm0.5\%$ within this range, which fully includes the fields measured in this work (on the order of $0.1$ mT). In addition, although the manufacturer specifies an operating temperature range from $-40^{\circ}$C to $+150^{\circ}$C, we verified that the sensors maintain good linearity even at cryogenic temperatures. Two initial tests were performed: first, by immersing a sensor in liquid nitrogen, and second, by placing it in a 4 K cryostat. In both cases, proper operation was confirmed. Finally, as demonstrated by the results presented in this work, these sensors also operate down to millikelvin temperatures, providing excellent linearity for the measured field values. This observation is particularly relevant for applications of TMR-based sensors in cryogenic or superconducting environments, where reliable magnetic field measurements are required at very low temperatures. Our results demonstrate that this commercial sensor operates reliably under such conditions.

\subsection{Discussion on materials and manufacturing}


Although pure aluminum (a low-temperature superconductor, LTS) was used in this work, similar results should be achievable with high-temperature superconductors (HTS). 
However, the practical realization of the system could become more challenging due to the brittleness and fabrication constraints typically associated with HTS materials. Alternatively, superconducting tapes could be employed as surface layers instead of bulk elements, as long as superconductivity is ensured.

From a manufacturing perspective, alternative materials and fabrication techniques could be explored for future implementations. For instance, lost-wax (cire perdue) casting could offer certain advantages over the use of machining for producing more complex geometries, as well as for achieving higher critical temperatures and critical fields, while additive manufacturing approaches---such as laser sintering of tantalum or other superconducting metals---could enable high-precision fabrication of intricate components compatible with cryogenic operation.

Regarding the ferromagnetic components---which could also be implemented in the form of foils or thin sheets---some hysteresis effects may arise in the cast-iron end caps, although no significant influence was observed under the quasi-static, low-field conditions of the present experiment. Cast iron was, in fact, chosen because it combines very low coercivity with high permeability and linear behavior within the applied field range (well below magnetic saturation). Other soft ferromagnetic materials, such as mu-metal or permalloy, could also serve as suitable candidates to achieve equivalent properties for the IMP surfaces.

\section{Numerical methods}

Numerical simulations were performed using COMSOL Multiphysics 6.2 with the Magnetic Fields (mf) interface, which solves the equations $\nabla\times\mathbf{H}=\mathbf{J}$, $\mathbf{B}=\nabla\times\mathbf{A}$, and $\mathbf{J}=\sigma\mathbf{E}$, with boundary conditions $\mathbf{n}\times\mathbf{A}=0$ at the borders of the simulation domain. 
The simulations were carried out in the stationary state, using either three-dimensional (3D), two-dimensional (2D), or 2D axisymmetric spatial representations as appropriate. 
The simulation mesh was defined with a maximum element size of 0.05 mm, a minimum element size of 0.01 mm, a maximum element growth rate of 1.1, a curvature factor of 0.2, and a resolution of narrow regions of 1. For material properties, a relative permeability of $1\cdot10^5$ was assigned to IMP media and $1\cdot10^{-5}$ to ZMP materials.
The total calculation domain in cylindrical axisymmetry, used for simulations corresponding to the experiment, was a rectangle of 350 mm $\times$ 500 mm, with the region of interest approximately 60 mm $\times$ 160 mm, containing the study region of 22.5 mm $\times$ 92 mm.